# Observation of Multi-Electromagnetically Induced Transparency in V-type Rubidium Atoms

**Kang Ying [1], Yueping Niu [2, *], Yihong Qi [2], Dijun Chen [1, †], Haiwen Cai [1], Ronghui Qu [1], Shangqing Gong [2, ‡]**

[1] *Shanghai Institute of Optics and Fine Mechanics, Chinese Academy of Sciences, Shanghai 201800, China*

[2] *Department of Physics, East China University of Science and Technology, Shanghai 200237, China*

*niuyp@ecust.edu.cn

†djchen@siom.ac.cn

‡sqgong@ecust.edu.cn

**Abstract.** A detailed experimental investigation and theoretical analysis have been made in the V-type $^{85}Rb$ atomic medium. Seven induced transparency windows, including a central double-peak-structure, have been observed experimentally when a coupling field and a probe field are applied into the ground and first excited states. By taking into account the hyperfine splitting of the excited state, our theoretical analysis gives good explanation for the observed phenomena.

## 1. Introduction

Electromagnetically induced transparency (EIT) is a quantum interference phenomenon occurring when two electromagnetic fields resonantly excite two different transitions sharing a common state. When a strong coupling laser is present, a weak probe laser can propagate through an opaque atomic medium. Since its first demonstration [1, 2], EIT effect has attracted considerable attention because of its potential applications in many fields, such as group velocity control [3-10], nonlinear susceptibility modulation [11-19], quantum information processing [20-22], quantum metrology [23], Rydberg atomic-state detection [24,25], laser frequency stabilization [26, 27], etc.

Many studies of EIT are based on ideal three-level models in Lambda-, V- or Ladder types. For example, Fulton *et al.* have made a composite theoretical and experimental study into the viability of Lambda-, V- and Ladder-type systems within rubidium for the observation of EITs. They pointed out that more complicated factors, including optical pumping effect, residual Doppler broaden effect, coupling field saturation effect, etc., exist in practical experiments than is suggested by ideal three-level models [28]. Up to now, many EIT researches have been concentrating on the V-type three-level system. Welch *et al.* conducted a study of V-type EIT in sodium experimentally for the purpose of demonstrating EIT as a quantum interference effect while eliminating the population trapping effect [29]. Recently, V-type EIT experiment has also been extended to $Na_2$ molecular system [30]. But actually, atoms or molecules often have complicated hyperfine level structures. In a practical experiment, more than three levels will interact with the incident fields. Zhao *et al.* measured two absorption dips experimentally in the V-type configuration which corresponds to the two hyperfine levels of the $6^2P_{3/2}$ $F'=4$ and $F'=5$ [31]. Vdovic *et al.* observed one EIT in $^{85}Rb$ isotope including hyperfine splitting of excited states $5^2P_{3/2}$ and $6^2P_{3/2}$. They discussed the optical pumping and coupling field saturation effect which mask the EIT signal in the V-type configuration in detail [32]. In this paper, we will do a detailed experimental investigation on the EIT in V-type $^{85}Rb$ atoms when both the coupling and probe fields are applied into the ground and first excited states. Totally seven induced transparency windows with a central double-peak-structure are observed at the vapor temperature of $70^0$C when the optical pumping effect is suppressed. Further, the EIT effect is distinguished from the coupling saturation effect. With the change of the coupling field frequency, the position of the multi-EIT windows could be moved accordingly while the frequency separations between them are constant. Theoretical analysis is given which shows good agreement with the experimental results.

## 2. Experimental Results and theoretical analysis

The V-type system we used for experimental study is the $D_2$ line (780 nm) of rubidium 85. A strong coupling laser and a weak probe laser are both applied between the ground state $5^2S_{1/2}$ and the excited state $5^2P_{3/2}$, as Fig. 1 shows. A diagram of the experimental setup is shown in Fig. 2. Both of the probe and coupling lasers are single-mode tunable external cavity diode lasers (ELDL) (New Focus TLB-6900), which has a half line-width of about 300 kHz._The half-wave plate 1 (HWP1) and beam splitter 1 (PB1) are used to attenuate the power of probe laser to below 50 $\mu$W to avoid the saturated absorption of the $^{85}Rb$ atoms and self- focusing effect. The half-wave plate 2 (HWP2) and beam splitter 3 (PB3) are used to attenuate the power of coupling laser to below 2 mW for the V- type system to weaken the power broadening effect. The probe and coupling beams are brought together by the beam splitter 2 (PB2) and co-propagating configuration is used to reduce the effect of Doppler broadening. The probe and coupling beams are orthogonally polarized when they enter the 75 mm-long Rb cell through PB2. The coupling beam is rejected by the beam splitter 4 (PB4) before reaching the detector. The Rb vapor cell is kept in a magnetic shielding structure to eliminate the earth magnetic field effect.

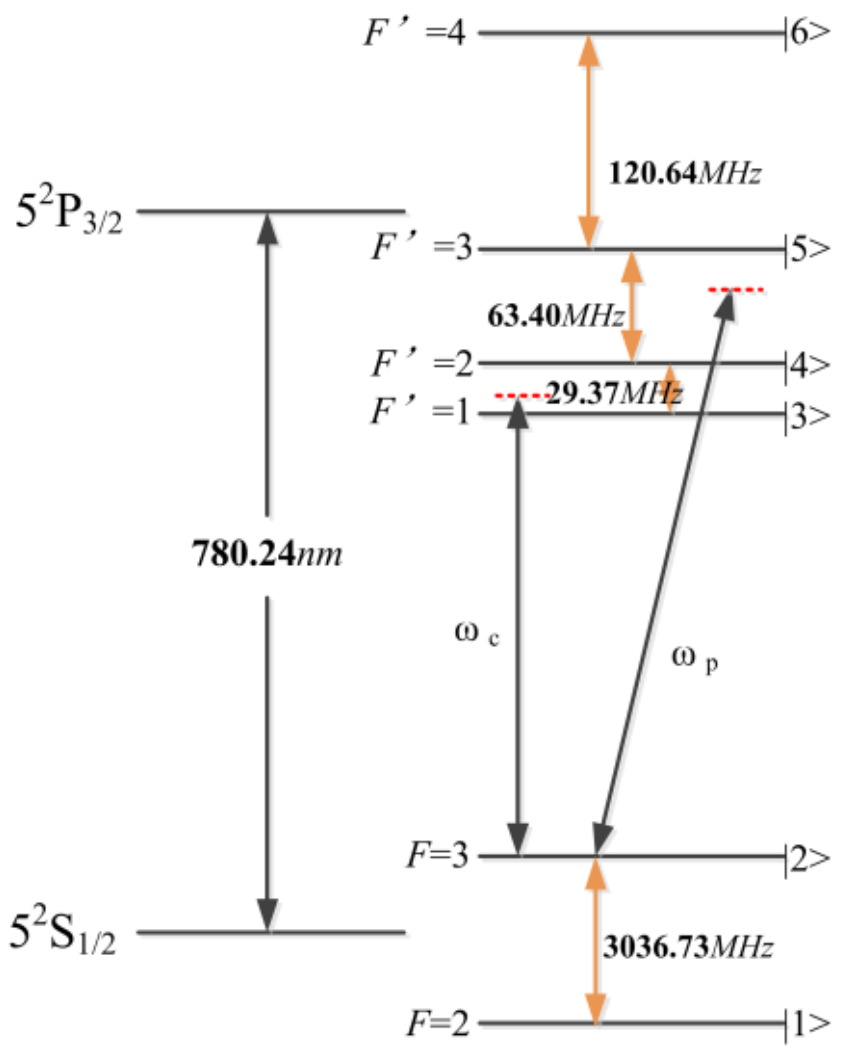

**Figure 1.** The energy level configuration of the $^{85}Rb\ D_2$ line for our V-type EIT experiment

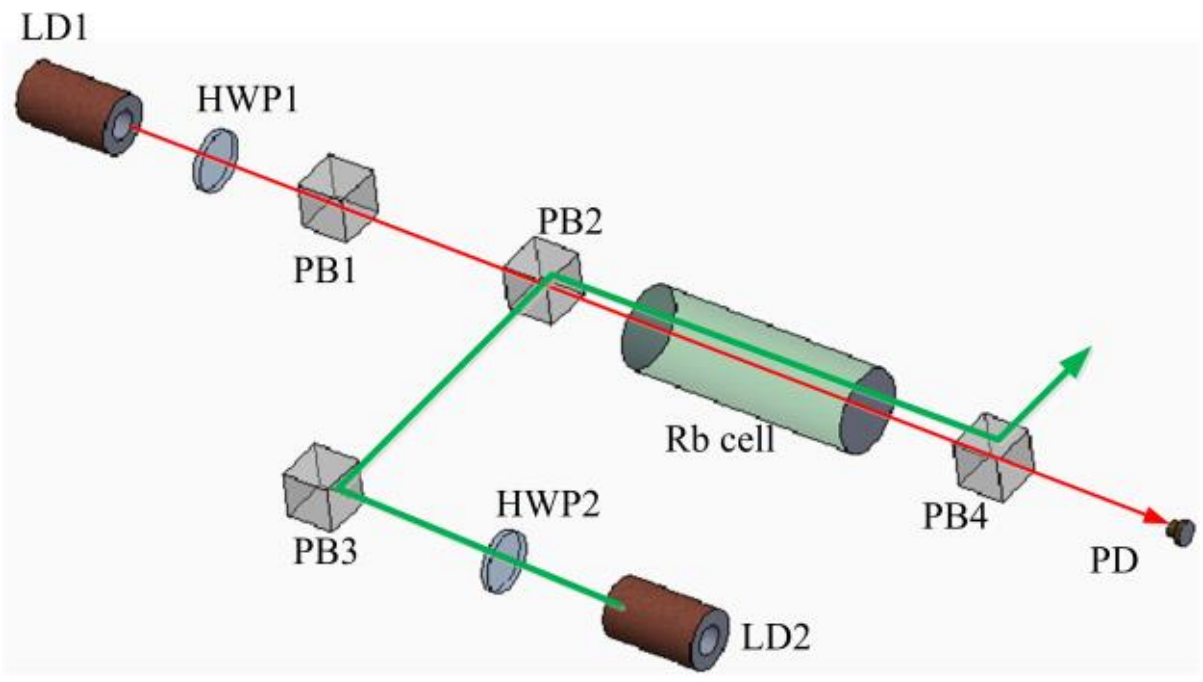

**Figure 2.** Schematic diagram of the experimental setup:
LD1 and LD2: probe and coupling lasers; PB1-PB4: polarizing cubic beam splitters;
HWP1-HWP2: half-wave plates; PD: photodiode detector.

As is well known, the coupling field will share the initial level with the weak probe field in the V-type configuration and cause a strong optical pumping effect that coincides with the EIT resonance. As is reported in [32], this mechanism can be suppressed by heating the Rb atoms. In this way, the thermalization of the ground-state hyperfine level populations will be faster than one optical pumping cycle. In our experiment, the vapor temperature is controlled through a specially designed copper tubing surrounded by the heating wire. Figure 3 shows the probe field transmission versus the detuning of the probe field with and without the coupling field when the vapor temperature goes up to $70^0$C. In the case that the coupling field is turned off, two broad absorption windows are recorded. They correspond to the hyperfine splitting of the ground state, which is 3036 MHz for $^{85}$Rb. When the coupling field is turned on, we tune it to resonant with the transition of $5^2S_{1/2}$(F=3) -> $5^2P_{3/2}$(F'=3). Then, Lambda-type EIT and V-type EIT form in the two broad absorption windows. As Fig. 3 shows, the lack of increase in the whole absorption of $5^2S_{1/2}$(F=2) -> $5^2P_{3/2}$(F'=1, 2, 3) indicates that the optical pumping effect has been suppressed at the present temperature of $70^0$C. In order to see the spectra clearly, we zoom it in Fig. 4. From this figure, first we could see that seven EIT windows occur. We can measure the frequency separations between these windows using the tunable coefficient of the probe laser. The frequency separations between the seven EIT windows are 63.2*MHz*, 57.5*MHz*, 63.2*MHz*, 63.1*MHz*, 57.2*MHz*, 63.6*MHz*, respectively. Second, we find that the central EIT signal has an obvious double-peak structure, which is very different from the other six EIT windows. Moreover, when we tune the frequency of the coupling laser to a higher 75*MHz*, all the seven EIT windows moved 75*MHz* accordingly while the separations between them remain constant, just as Fig. 5 shows.

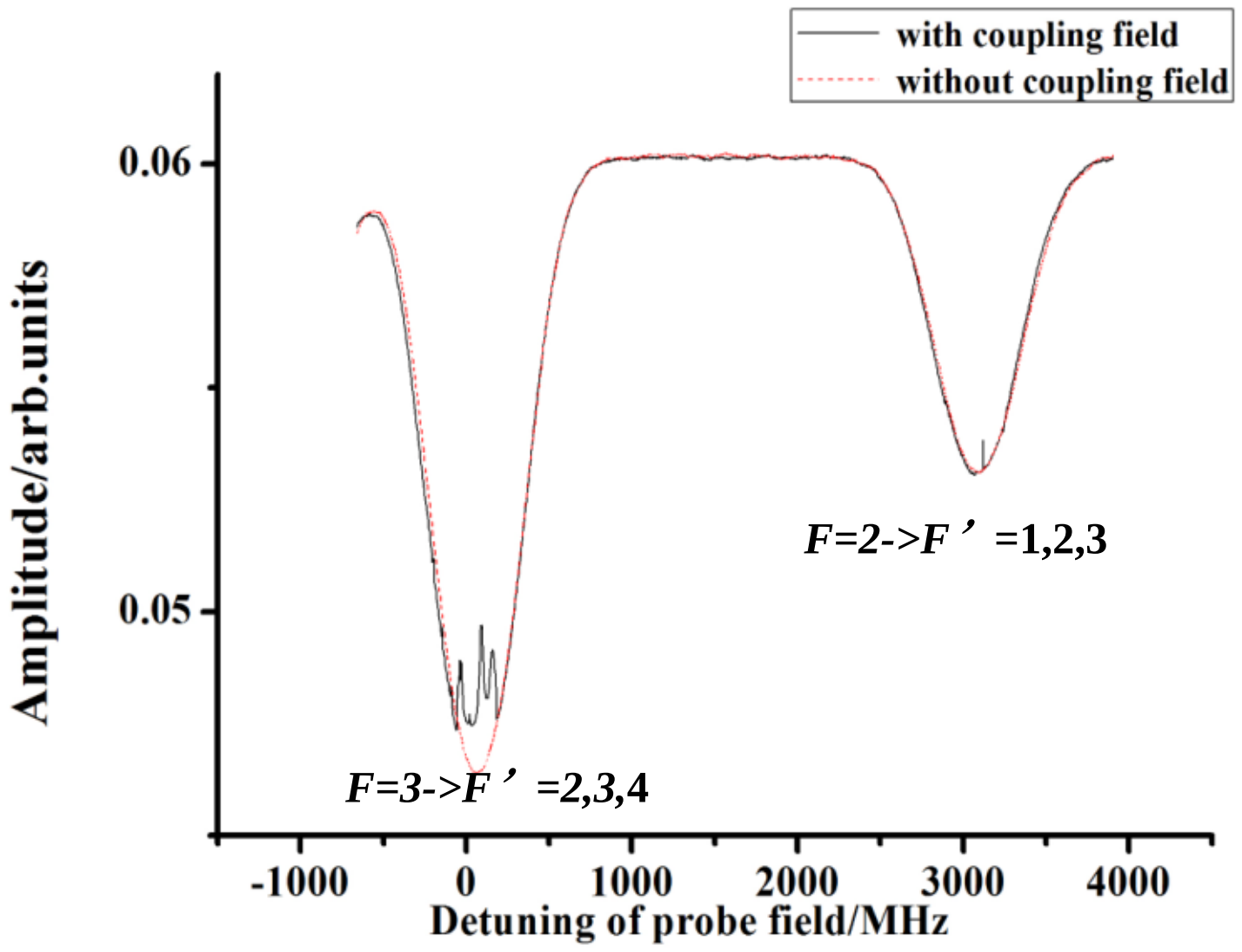

**Figure 3.** Experimental trace of probe field absorption with and without the coupling field (1.95*mW* coupling field, 45uW probe field) at the vapor temperature of about $70^0$C.

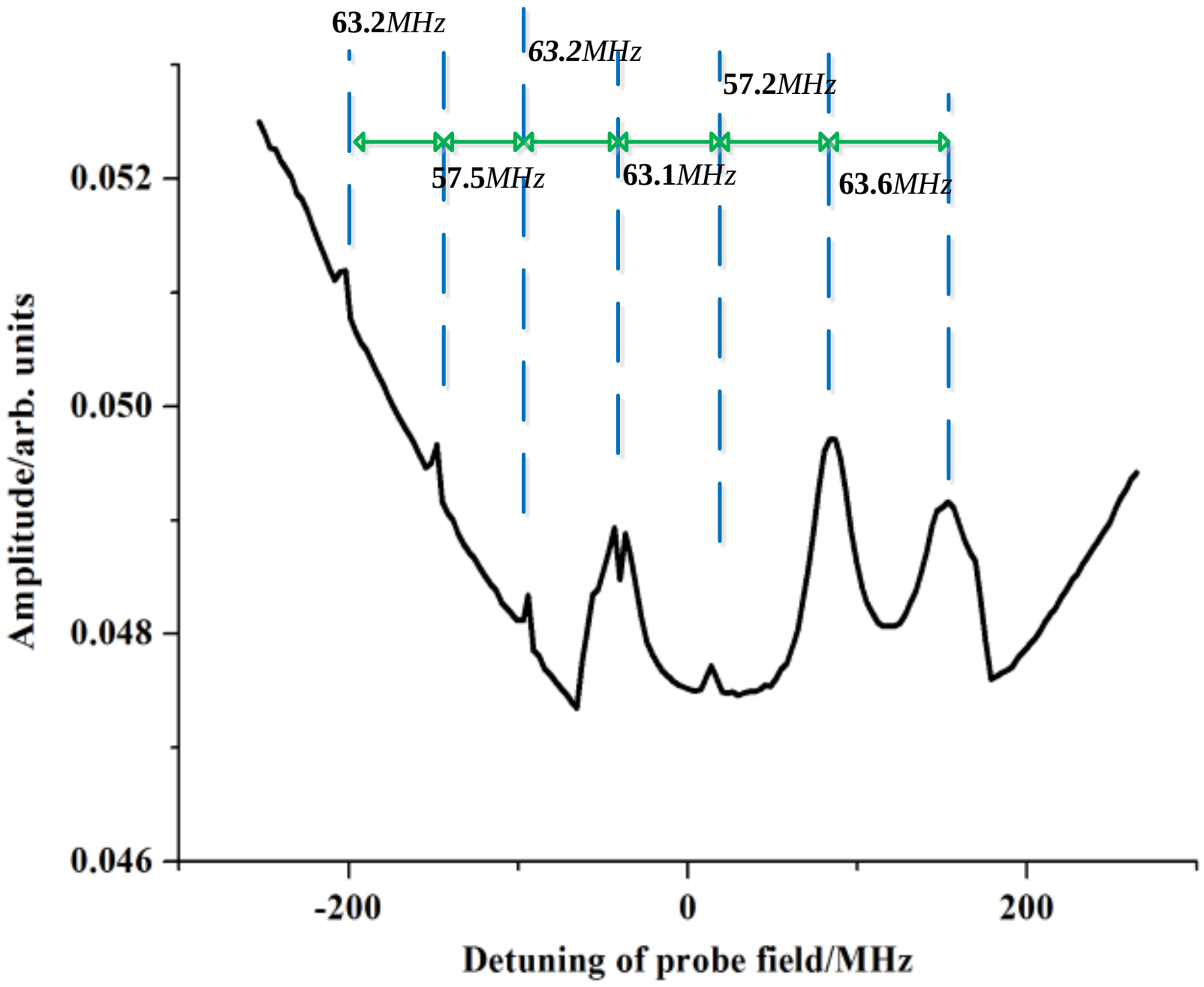

**Figure 4.** Experimental trace of the seven EITs in the V-type $^{85}Rb$ medium using 1.95*mW* coupling field at vapor temperature of $70^0$C.

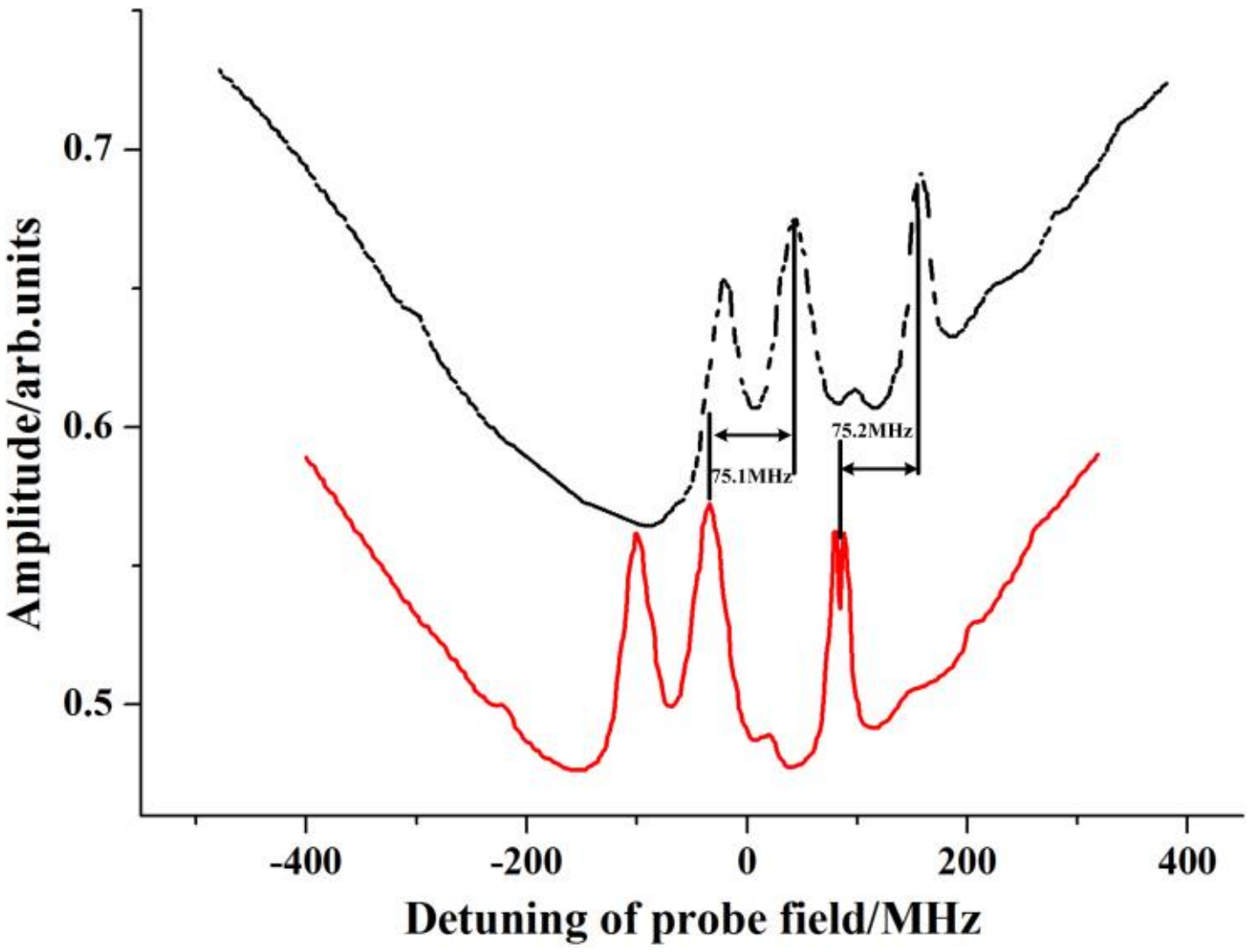

**Figure 5.** Moving of the seven EIT windows with the tuning of the coupling laser frequency: The solid line indicates the initial experimental trace of V-type EIT; the dotted line indicates the experimental trace when tuning the frequency of the coupling laser to a higher 75*MHz*.

In order to explain the above experimental results, theoretical analysis of the V-type system is needed. As we know, there would be an EIT window when scanning the probe laser field to meet the two-photon resonance condition. If we define $\Delta_{pi} = \omega_p - \omega_{i2}$, $\Delta_{cj} = \omega_c - \omega_{j2}$ (see the atomic energy structure of figure 1, $\omega_{i,j\,2}$ is the frequency separation between the level |*i*,*j*> and |2>, *i*, *j*=4, 5, 6) as the frequency detuning for the probe and coupling field respectively, then EIT occurs at the position of $\Delta_{pi} = \Delta_{cj}$. According to the transition rule, each transition from $F = 3 \rightarrow F' = 2, 3, 4$ is possible when the probe and coupling laser fields are applied between the ground and the first exited states. As a result, there would appear multi-EIT windows in the V-type system of $^{85}Rb$ which is quite different from the Lambda-type case where only one EIT window occurs. Following is a detailed analysis.

For *j*=4, as seen in figure 6(a), there would be three values of $\omega_p$ which could meet the two-photon resonance condition, i.e., $\Delta_{p4} = \Delta_{c4}$, $\Delta_{p5} = \Delta_{c4}$ and $\Delta_{p6} = \Delta_{c4}$. The frequency separations between these three EIT windows are $\Delta\omega_{54}$ and $\Delta\omega_{64}$ in accordance with the frequency separations between the hyperfine splitting levels |4>, |5> and |6>. For the other two cases of *j*=5 and *j*=6, EIT forms when $\Delta_{pi} = \Delta_{c5}$ and $\Delta_{pi} = \Delta_{c6}$, as figure 6(b) and 6(c) shows. Therefore, nine EIT windows should present totally. However, in each case of figure 6, one EIT occurs at $\Delta_{pi} = \Delta_{cj}$ when $i = j$. Then, these three EITs for $i = j$ displayed the common one. As a result, seven EIT windows appear when scanning the probe laser frequency. On the other hand, the position of EIT for $i = j$ is just the central one and it has a double-peak structure. As we know, when $i = j$, it means that the probe and coupling field interact with the same ground and excited hyperfine splitting level. We have noted that similar spectra have been widely investigated in other experiments and different physical mechanisms are put forward for the explanations [33-38]. For the detailed investigation of the double-peak-structure spectra, specific atomic transition and polarization of the coupling and probe lasers are needed and is now under study.

Furthermore, according to the above analysis, the frequency separations between these seven EIT windows should be $\Delta\omega_{54}$, $\Delta\omega_{65}$ - $\Delta\omega_{54}$, $\Delta\omega_{54}$, $\Delta\omega_{54}$, $\Delta\omega_{65}$ - $\Delta\omega_{54}$ and $\Delta\omega_{54}$, just as table 1 shows. This also indicates that the frequency separations between the EIT windows will not change with the applied field parameters. But, these seven EIT windows could make a red-shift or a blue-shift in accordance with the detuning of the coupling laser field. Comparing the experimental results with the theoretical analysis, we can see that they are in good consistency.

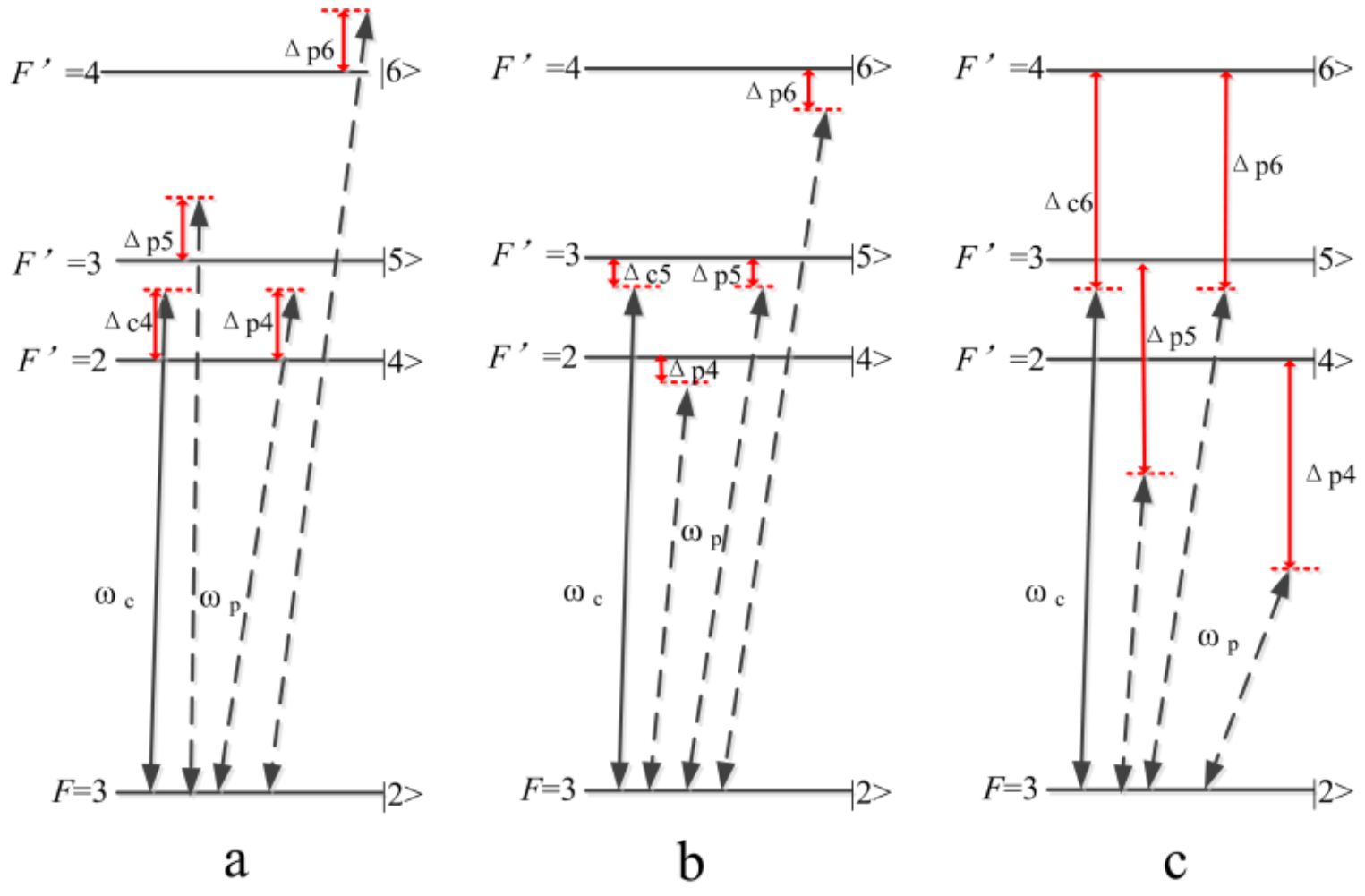

**Figure 6.** Theoretical analysis for the seven EIT windows when scanning the probe laser field to meet the two-photon resonance condition. The black solid lines indicate the transition induced by coupling laser field; the dotted lines indicate the transition induced by probe laser field.

**Table 1.** Relative position of EIT windows in the V-type system in $^{85}$Rb

| | | Observed value (*MHz*) | Calculated value (*MHz*) |
|---|---|---|---|
| 1 | 0 | 0 | 0 |
| 2 | $\Delta\omega_{54}$ | 63.2 | 63.4 |
| 3 | $\Delta\omega_{65}$ - $\Delta\omega_{54}$ | 57.5 | 57.2 |
| 4 | $\Delta\omega_{54}$ | 63.2 | 63.4 |
| 5 | $\Delta\omega_{54}$ | 63.1 | 63.4 |
| 6 | $\Delta\omega_{65}$ - $\Delta\omega_{54}$ | 57.2 | 57.2 |
| 7 | $\Delta\omega_{54}$ | 63.6 | 63.4 |

It is valuable to indicate two important factors for the observation of theoretical predicted multi-EIT windows in the V-type system. One factor is the separations of the hyperfine splitting levels. Just as we analyzed above, the frequency separations between the multi-EIT windows depend on the separations of the hyperfine splitting levels. In the case of very small hyperfine splitting, the separations between the multi-EIT windows are also very small and then the multi-EIT windows become undistinguishable [32]. While in the opposite case, very large hyperfine splitting will result in the far-off resonance of the coupling field. Then the interaction weakens and hence the corresponding EIT effect becomes negligible [31]. The other factor is the coupling field power used in the experiment. In the case of weak coupling field, the transparency is weak since the EIT is not fully developed. While in the opposite case, a quite strong coupling field will cause considerable power broadening effect in the V-type configuration. The transparency windows will be broadened and hence the adjacent transmission peaks will overlap. In view of this, we choose the $D_2$ line of $^{85}Rb$ as the experimental medium whose upper hyperfine splitting levels are suitable for the observation of seven EITs with a ~2$mW$ coupling laser.

In our experiment, except for the suppressed optical pumping effect, the coupling field saturation is also a very important incoherent effect that along with the EIT. It can never be separated experimentally from the EIT signal in the V-type configuration. But as is addressed in [32], the coupling field saturation is an incoherent effect which does not depend on the unlinked states coherence, while the coherent EIT is strongly dependent on the unlinked states coherence. As the unlinked states coherence decay increases, the EIT would be reduced and the width of EIT windows would be broadened clearly. Based on this principle, we carry out theoretically simulation utilizing the simplified three-level V-type scheme (see the energy levels in Fig.7). According to the experimental parameters, the calculation was performed for coupling field Rabi frequency of 100MHz (It is resonant with F'=3). In Fig. 7, we present the calculated probe absorption line profiles for different values of the unlinked states coherence decay rates. It is clearly that as the coherence decay rate increases from 5MHz to 50MHz, the width of the transmission profile is increased from 25MHz to 55MHz. This 55MHz linewidth just coincides to the linewidth of the seventh transparency window on the right in Fig. 4. This verifies that the coherent EIT effect exists in the combined spectra of Fig. 4.

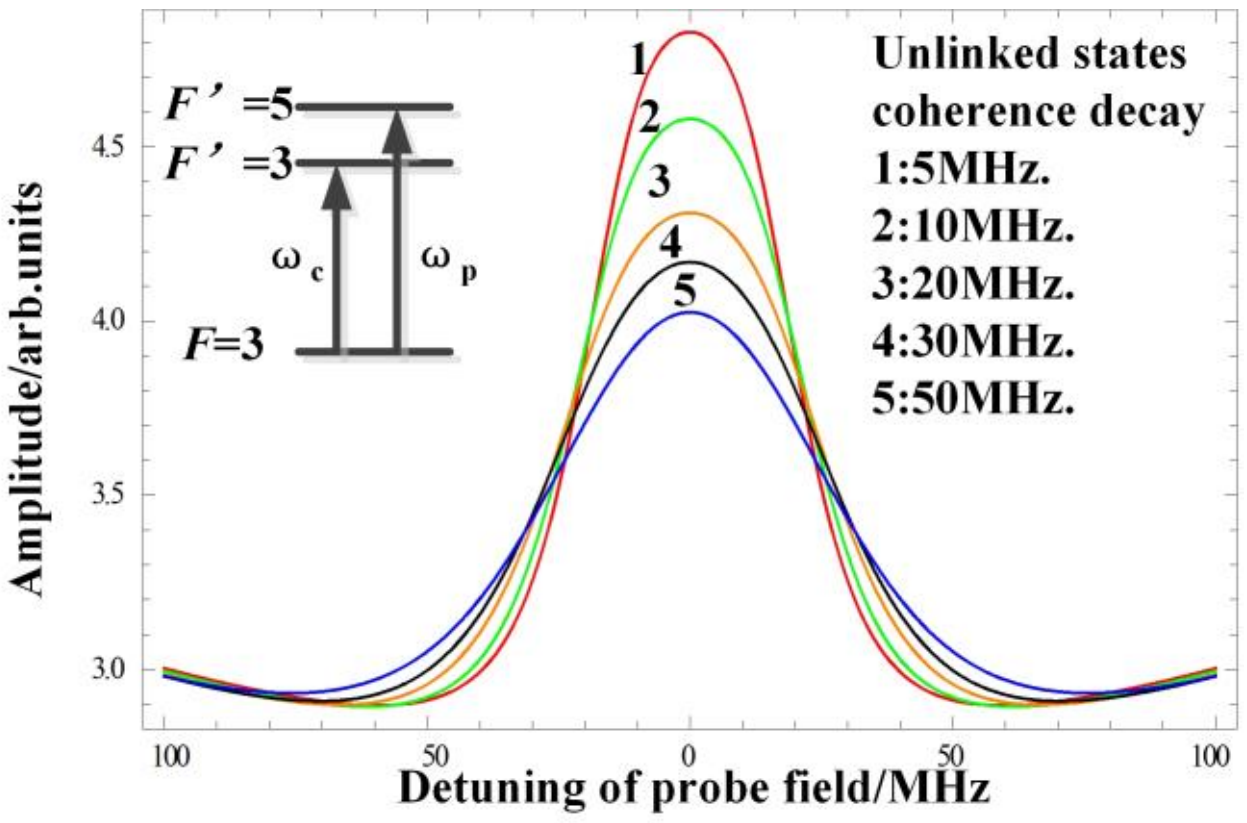

**Figure 7.** Calculated probe profiles for different values of unlinked states coherence decay rates.

## 3. Conclusions

In summary, we have made a detailed study of the multi-EIT in the V-type $^{85}Rb$ medium experimentally and theoretically. Totally seven induced transparency windows have been observed, of which their separations are related to the structure of the upper hyperfine splitting level. At the Rb vapor temperature of 70$^0$C, the optical pumping effect is suppressed. Also, the EIT effect is distinguished from the coupling saturation effect via comparing the numerical simulation and experimental result. Two important factors for observing all the seven induced transparency windows have been addressed. Our theoretical analysis gives good explanations to the observed experimental phenomena.

**Acknowledgement:**

This work was supported by the National Natural Science Foundation of China (Grant Nos. 60978013, 61108028 and 61178031) and Shanghai Rising-Star Program of Grant No. 11QA1407400.